\documentclass[showpacs,preprintnumbers,superscriptaddress]{revtex4}
\usepackage{CJK}
\usepackage{amsmath,amssymb,graphicx,bm}
\begin{document}

\title{Accretion of the Vlasov gas onto a Schwarzschild-like black hole}

\author{Ziqiang Cai}
\affiliation{College of Physics Science and Technology, Hebei University, Baoding 071002, China}

\author{Rong-Jia Yang \footnote{Corresponding author}}
\email{yangrongjia@tsinghua.org.cn}
\affiliation{College of Physics Science and Technology, Hebei University, Baoding 071002, China}
\affiliation{Hebei Key Lab of Optic-Electronic Information and Materials, Hebei University, Baoding 071002, China}
\affiliation{National-Local Joint Engineering Laboratory of New Energy Photoelectric Devices, Hebei University, Baoding 071002, China}
\affiliation{Key Laboratory of High-pricision Computation and Application of Quantum Field Theory of Hebei Province, Hebei University, Baoding 071002, China}

\begin{abstract}
We consider spherical steady accretion of the relativistic Vlasov gas onto a Schwarzschild-like black hole. We determine the expressions for the particle current density and accretion rate and present the limiting expressions for the mass accretion rate at high and low temperature. The results show that the parameter characterizing the breaking of Lorentz symmetry can affects the radial component of the particle current density and the mass accretion rate.
\end{abstract}

\maketitle

\section{Introduction}
Accretion of matter onto a massive object is one of the most common processes in astrophysics \cite{Yuan:2014gma}. The pioneers' works on accretion date back to early papers by Hoyle, Lyttleton, and Bondi \cite{hoyle1939effect, 1940Obs6339L, Bondi:1944jm}. Bondi first derived Newtonian solutions representing spherically symmetric accretion of perfect fluid in the Keplerian gravitational potential \cite{Bondi:1952ni}. Michel considered spherical accretion of the perfect fluid onto the Schwarzschild black hole, which generalized the Bondi's model to General Relativity \cite{michel1972accretion}. Since then accretion has been extensively studied by numerous works in the literature, see for example \cite{begelman1978accretion, Malec:1999dd, Babichev:2004yx, Babichev:2010kj, Rodrigues:2016uor, Contreras:2018gct, Abbas:2018ygc, Zheng:2019mem, Yang:2020bpj, UmarFarooq:2020aum, Nozari:2020swx, Panotopoulos:2021ezt, Iftikhar:2020ykp, Gao:2008jv, John:2013bqa, Jiao:2016iwp, Ganguly:2014cqa, Mach:2013fsa, Kremer:2020yfg, Tejeda:2019lie, Feng:2022bst}. These works mainly focused on critical points, flow parameters, accretion rate, and so on. In \cite{Yang:2015sfa}, accretion onto a renormalization-group-improved Schwarzschild black hole was used to test the asymptotically safe scenario. While in \cite{Yang:2018zef}, accretion on a Schwarzschild-like black hole was investigated to test the Lorentz symmetry. In \cite{Jamil:2008bc, Yang:2019qru}, the conditions for accretion give limits on the ratio of mass to charge.

In general, numerical analysis are needed to handle nonspherical accretion for either relativistic or Newtonian flow \cite{Papadopoulos:1998up, Font:1998sc, Zanotti:2011mb, Lora-Clavijo:2015hqa}. So it is important to find analytical solutions for accreting processes. Exact fully relativistic solutions were obtained in \cite{Petrich:1988zz} describing for matter with adiabatic equation of state accreted onto a moving Schwarzschild or Kerr black hole. Based on this work, analytic solutions were found for accretion onto a moving Kerr-Newman black hole \cite{Babichev:2008dy}, onto a moving Reissner-Nordstr{\"o}m \cite{Jiao:2016uiv}, and onto a moving charged dilaton black hole \cite{Yang:2021opo}. In \cite{Liu:2009ts, Zhao:2018ani}, exact solutions were derived for shells accreted onto a Schwarzschild black hole. Recently, exact solutions were presented for accretion of collisionless Vlasov gas onto a moving Schwarzschild black hole \cite{Mach:2021zqe, Mach:2020wtm} or onto a Kerr black hole \cite{Cieslik:2022wok}, which bases on the Hamiltonian formalism developed in \cite{Rioseco:2016jwc}. This method makes it is possible for analysing more complex flows on a fixed background, but because of the computational complexity, it is difficult to obtain the exact solution of mass accretion rate for more complicated black hole (see for example a Reissner-Nordstr{\"o}m black hole \cite{Cieslik:2020ibk}).

Lorentz symmetry breaking (LSB) is an interesting topic in Physics. Many LSB models have been constructed in the literatures.
The bumblebee model is a known gravity model that extends the standard formalism of general relativity. Under a suitable potential, the bumblebee vector field acquires a nonvanishing vacuum expectation value which triggers a spontaneous LSB \cite{Kostelecky:1989jp, Kostelecky:2003fs, Bluhm:2004ep, Casana:2017jkc}. A static and spherically symmetric Schwarzschild-like black hole was obtained in bumblebee gravity \cite{Casana:2017jkc}. In \cite{Yang:2018zef}, the effects of the spontaneous LSB were investigated in the process of accretion onto the Schwarzschild-like black hole. Here we will use the Hamiltonian formalism method \cite{Rioseco:2016jwc, Mach:2021zqe, Mach:2020wtm, Cieslik:2020ibk} to consider the accretion of Vlasov gas onto the Schwarzschild-like black hole, and investigate whether we can obtain a compact formula for the mass accretion rate? If can, what effect will the parameter characterizing the effect of LSB have on the particle current density? It is also interesting to compare the results obtained here with those in \cite{Yang:2018zef}.

The order of this paper is as follows. In Section II, we will briefly review the bumblebee gravity and its Schwarzschild-like solution. In Section III, we will give the basic equations satisfied by the gas in the Schwarzschild-like background. In Section IV, we will discuss the distribution functions obeyed by the gas. In Section V, we will present expressions for the particle current density and the mass accretion rate. Finally, we will briefly summarize and discuss our results in Section VI.

\section{Bumblebee gravity model}
The bumblebee gravity models are the simplest examples of field theories with spontaneous Lorentz and diffeomorphism violations. In these scenarios, the spontaneous LSB is induced by a potential whose functional form possesses a minimum which breaks the $U(1)$ symmetry. For a single bumblebee field $B_\mu$ coupled to gravity and
matter, the action can be written as \cite{Casana:2017jkc}
\begin{equation}
\label{01}
S_B=\int d^4 x \mathcal{L}_B,
\end{equation}
with
\begin{equation}
\begin{aligned}
\label{02}
\mathcal{L}_B=\frac{e}{2 \kappa} R+\frac{e}{2 \kappa} \xi B^\mu B^\nu R_{\mu \nu}-\frac{1}{4} e B_{\mu \nu} B^{\mu \nu}-e V\left(B^\mu\right)+\mathcal{L}_{\mathrm{M}},
\end{aligned}
\end{equation}
where $e\equiv\sqrt{-g}$ is the determinant of the vierbein and $\xi$ is the real coupling constant which controls the nonminimal gravity-bumblebee interaction.  We takes the unit as $G=c=1$ throughout this paper. The bumblebee field strength is defined as
\begin{equation}
B_{\mu \nu}=\partial_\mu B_\nu-\partial_\mu B_\nu.
\end{equation}
From the Lagrangian density (\ref{02}), yields the modified Einstein equations
\begin{equation}
\label{03}
R_{\mu \nu}-\frac{1}{2} R g_{\mu \nu}=\kappa T_{\mu \nu},
\end{equation}
where $T_{\mu \nu}$ is the total energy-momentum tensor which reads
\begin{equation}
T_{\mu \nu}=T_{\mu \nu}^{\mathrm{M}}+T_{\mu \nu}^B,
\end{equation}
with
\begin{equation}
\begin{aligned}
T_{\mu \nu}^B=&-B_{\mu \alpha} B_\nu^\alpha-\frac{1}{4} B_{\alpha \beta} B^{\alpha \beta} g_{\mu \nu}-V g_{\mu \nu}+2 V^{\prime} B_\mu B_\nu \\
&+\frac{\xi}{\kappa}\left[\frac{1}{2} B^\alpha B^\beta R_{\alpha \beta} g_{\mu \nu}-B_\mu B^\alpha R_{\alpha \nu}-B_\nu B^\alpha R_{\alpha \mu}\right.\\
&+\frac{1}{2} \nabla_\alpha \nabla_\mu\left(B^\alpha B_\nu\right)+\frac{1}{2} \nabla_\alpha \nabla_\nu\left(B^\alpha B_\mu\right) \\
&\left.-\frac{1}{2} \nabla^2\left(B_\mu B_\nu\right)-\frac{1}{2} g_{\mu \nu} \nabla_\alpha \nabla_\beta\left(B^\alpha B^\beta\right)\right],
\end{aligned}
\end{equation}
where the prime means differentiation with respect to the argument. The equation of motion for the bumblebee field from (\ref{02}) is given by
\begin{equation}
\nabla^{\mu}B_{\mu\nu}=J_\nu^M+J_\nu^B,
\end{equation}
where $J_\nu^M$ is the matter current and $J_\nu^B$ is the bumblebee field current which takes the form
\begin{equation}
J_\nu^B=2 V^{\prime} B_\nu-\frac{\xi}{\kappa} B^\mu R_{\mu \nu}.
\end{equation}
A static and spherically symmetric solution in bumblebee gravity (\ref{02}), called Schwarzschild-like black hole, was obtained in \cite{Casana:2017jkc}, its geometry is given by the following line element
\begin{eqnarray}
\label{12}
ds^2=-N(r)d\overline{t}^{2}+(1+l)\frac{1}{N(r)}d{r}^{2}+r^{2}(d\theta^{2}+\sin^{2}\theta{d\varphi^{2}}),
\end{eqnarray}
where $N=1-2M/r$ with $M$ the mass of the black hole and $l$ is a constant characterizing the LSB. Since $g_{11}>0$, we have theoretical constraints on the parameter $l$: $l>-1$. In \cite{Casana:2017jkc}, an upper-bound for $l$ was found: $l\leq 10^{-13}$. The horizon of black hole is at $r_{H}=2M$.

Here we will consider spherical steady accretion of the relativistic Vlasov gas onto the Schwarzschild-like black hole (\ref{12}) by using the Hamiltonian formalism method \cite{Rioseco:2016jwc, Mach:2021zqe, Mach:2020wtm, Cieslik:2020ibk} to investigate the effect of parameter characterizing the LSB on accretion rate and accretion flow.

\section{Vlasov equation in a Schwarzschild-like background}
In this Section, we will briefly review the Hamiltonian form of the geodesic motion, the Vlasov equation, Horizon-penetrating coordinates, and action-angle variables, which were discussed in detail in \cite{Rioseco:2016jwc}.

\subsection{Hamiltonian description of the geodesic motion}
Following \cite{Rioseco:2016jwc}, it is convenient to discuss particle motion in the Hamiltonian framework. We briefly review the Hamiltonian description of the geodesic motion. For a free particle moving along a time-like geodesic, the Hamiltonian can be chosen as
\begin{eqnarray}
\label{1}
H=\frac{1}{2}g^{\mu\nu}(x^{\alpha})p_{\mu}p_{\nu},
\end{eqnarray}
where $g^{\mu\nu}$ is the spacetime metric, $p^{\mu}=\mathrm{d}x^{\mu}/\mathrm{d}\tau$ is the momenta of the particle. $(x^{\mu},p_{\nu})$ are canonical variables related to the Hamiltonian. Since $H =\frac{1}{2}g^{\mu\nu}p_{\mu}p_{\nu}=-\frac{1}{2}m^{2}$ with $m$ the rest mass of the particle, consequently $\tau=s/m$ with $s$ the proper time. The equations of motion are given by
\begin{eqnarray}
\label{2}
\frac{\mathrm{d}x^{\mu}}{\mathrm{d}\tau}=\frac{\partial H}{\partial p_{\mu}}, \qquad \frac{\mathrm{d}p_{\nu}}{\mathrm{d}\tau}=-\frac{\partial H}{\partial x^{\nu}}.
\end{eqnarray}
From (\ref{2}), yields the standard geodesic equation
\begin{eqnarray}
\label{3}
\frac{\mathrm{d^{2}}x^{\mu}}{\mathrm{d}\tau^{2}}+\Gamma_{\alpha\beta}^{\mu}\frac{\mathrm{d}x^{\alpha}}{\mathrm{d}\tau}\frac{\mathrm{d}x^{\beta}}{\mathrm{d}\tau}=0,
\end{eqnarray}
where $\Gamma_{\alpha\beta}^{\mu}$ is the Christoffel symbols which is connected with the metric $g_{\mu\nu}$. The four-velocity $u^{\mu}=\mathrm{d}x^{\mu}/\mathrm{d}s$ satisfies the normalized conditions $g^{\mu\nu}u_{\mu}u_{\nu}=-1$.

\subsection{Vlasov equation}
Usually we use a probability function $f = f(x^{\mu},p_{\nu})$ \cite{Andreasson:2011ng} to describe the collisionless Vlasov gas. In the following we will look for the equation satisfied by the distribution function. The probability function should be invariant along a geodesic, so it must fulfil
\begin{eqnarray}
\label{4}
\frac{\mathrm{d}}{\mathrm{d}\tau}f(x^{\mu}(\tau),p_{\nu}(\tau))=0.
\end{eqnarray}
or equivalently
\begin{eqnarray}
\label{5}
0=\frac{\mathrm{d}x^{\mu}}{\mathrm{d}\tau}\frac{\partial {f}}{\partial x^{\mu}}+\frac{\mathrm{d}p_{\nu}}{\mathrm{d}\tau}\frac{\partial {f}}{\partial p_{\nu}}=\frac{\partial H}{\partial p_{\mu}}\frac{\partial {f}}{\partial x^{\mu}}-\frac{\partial H}{\partial x^{\nu}}\frac{\partial {f}}{\partial p_{\nu}}\equiv\{H,f\},
\end{eqnarray}
where $\{.,.\}$ is defined as the Poisson bracket. With (\ref{1}), the equation (\ref{5}) can be written as
\begin{eqnarray}
\label{6}
g^{\mu\nu}p_{\nu}\frac{\partial f}{\partial x^{\mu}}-\frac{1}{2}p_{\alpha}p_{\beta}\frac{\partial g^{\alpha\beta}}{\partial x^{\mu}}\frac{\partial f}{\partial p_{\mu}}=0,
\end{eqnarray}
which is usually called as the relativistic Liouville equation or the relativistic Vlasov equation. In the content above, the phase-space coordinates are chosen as ($x^{\mu}, p_{\nu}$) on the cotangent bundle. In the literature, it is usually to write the Vlasov equation (\ref{6}) in terms of coordinates ($x^{\mu}, p^{\nu}$) on the tangent bundle as
\begin{eqnarray}
\label{7}
p^{\mu}\frac{\partial f}{\partial x^{\mu}}-\Gamma^{\mu}_{\alpha\beta}p_{\alpha}p_{\beta}\frac{\partial f}{\partial p^{\mu}}=0.
\end{eqnarray}
For a collection of single-mass particles satisfying the mass shell condition, $g_{\mu\nu}p^{\mu}p^{\nu}=-m^2$, the Vlassov equation takes the form
\begin{eqnarray}
\label{8}
\frac{\partial f}{\partial t}+\frac{p^{i}}{p_{0}}\frac{\partial f}{\partial x^{i}}-\frac{1}{p^{0}}\Gamma^{i}_{\alpha\beta}p_{\alpha}p_{\beta}\frac{\partial f}{\partial p^{i}}=0,
\end{eqnarray}
which is probably the most common version of the Vlassov equation \cite{Andreasson:2011ng}. Obviously, the distribution function takes an important role in the Vlassov equation, it can be used to calculate some important observable quantities, such as the energy-momentum tensor $T_{\mu\nu}$ and the particle current density $J_{\mu}$. The latter can be expressed as an integral over momentum space
\begin{eqnarray}
\label{9}
J_{\mu}(x)=\int{p_{\mu}}f(x,p)\sqrt{-g}d^{4}p,
\end{eqnarray}
where $g=$det$[g^{\mu\nu}]$ is the determinant of the inverse metric tensor. The energy-momentum tensor is assumed as
\begin{eqnarray}
\label{10}
T_{\mu\nu}(x)=\int{p_{\mu}}p_{\nu}f(x,p)\sqrt{-g}d^{4}p.
\end{eqnarray}
With equation (\ref{6}), we can show that the particle current density satisfies the following conservation equation \cite{2013Relativistic}
\begin{eqnarray}
\label{11}
\nabla_{\mu}J^{\mu}=0,
\end{eqnarray}
where $\nabla_{\mu}$ is defined as the covariant derivative.

\subsection{Horizon-penetrating coordinates}
Since the coordinate system in (\ref{12}) is divergent at the horizon, we instead work in horizon-penetrating (Eddington-Finkelstein type) coordinates. Defining a
new time coordinate $t$
\begin{eqnarray}
\label{13}
t=\overline{t}+\int^{r}\left[\frac{1}{N(r)}-\eta(r)\right]dr,
\end{eqnarray}
where $\eta(r)$ is an arbitrary function. Then the metric (\ref{12}) takes the form
\begin{eqnarray}
\label{14}
ds^2=-Ndt^{2}+2\left(1-N\eta\right)dtdr+\left(\frac{l}{N}-N\eta^{2}+2\eta\right)dr^{2}+r^{2}\left(d\theta^{2}+\sin^{2}\theta{d\varphi^{2}}\right).
\end{eqnarray}
The corresponding contravariant components of metric are
\begin{eqnarray}
\label{15}
g^{tt}=-\frac{\frac{l}{N}-N\eta^{2}+2\eta}{1+l}, \qquad g^{tr}=\frac{1-N\eta}{1+l}, \qquad g^{rr}=\frac{N}{1+l}.
\end{eqnarray}
Note that
\begin{eqnarray}
\label{16}
(g^{tr})^{2}-g^{rr}g^{tt}=\frac{1}{1+l}.
\end{eqnarray}
Taking $\eta\equiv1$ is more convenient, also known as Eddington-Finkelstein coordinates.

\subsection{Action-angle variables}
For the Eddington-Finkelstein form of the Schwarzschild-like metric (\ref{14}), the Hamiltonian of a free particle can be expressed as
\begin{eqnarray}
\label{17}
H=\frac{1}{2}\left[g^{tt}(r)p_{t}^{2}+2g^{tr}(r)p_{t}p_{r}+g^{rr}(r)p_{r}^{2}+\frac{1}{r^{2}}\left(p_{\theta}^{2}+\frac{p_{\varphi}^{2}}{\sin^{2}\theta}\right)\right].
\end{eqnarray}
Because Hamiltonian $H$ does not depend on the $\tau$, and $H=-m^{2}/2$, meaning that $m$ is a constant of motion. Moreover, since $H$ dose not depend on $t$ and $\varphi$, the energy of moving particle $E\equiv -p_{t}$ and the azimutal angular momentum $\l_{z}\equiv p_{\varphi}$ are also constants of motion. Because of the spherical symmetry, there is another important conserved quantity, the total angular momentum, which is given by
\begin{eqnarray}
\label{18}
L=\sqrt{p_{\theta}^{2}+\frac{p_{\varphi}^{2}}{\sin^{2}\theta}}.
\end{eqnarray}
From the above mentioned conserved quantities, we can obtain
\begin{eqnarray}
\label{19}
g^{tt}(r)E^{2}-2g^{tr}(r)Ep_{r}+g^{rr}(r)p_{r}^{2}+\frac{L^{2}}{r^{2}}+m^{2}=0.
\end{eqnarray}
Solving Eq. (\ref{19}), we have the radial momenta as
\begin{eqnarray}
\label{20}	
p_{r}=\frac{(1-N\eta)E\pm\sqrt{(1+l)(E^{2}-\widetilde{U}_{l,m}(r))}}{N},
\end{eqnarray}
where we introduced the effective potential
\begin{eqnarray}
\label{21}
\widetilde{U}_{l,m}(r)=N\left(m^{2}+\frac{L^{2}}{r^{2}}\right).
\end{eqnarray}
From the expression (\ref{18}) of the total angular momenta $L$, we can get latitudinal momenta $p_{\theta}$ as
\begin{eqnarray}
\label{22}
p_{\theta}=\pm\sqrt{L^{2}-\frac{p_{\varphi}^{2}}{\sin^{2}\theta}}=\pm\sqrt{L^{2}-\frac{l_{z}^{2}}{\sin^{2}\theta}}.
\end{eqnarray}
We introduce a canonical transformation which can be taken as new momentum
\begin{eqnarray}
\label{23}
P_{0}=m=\sqrt{-g^{tt}(r)p_{t}^{2}-2g^{tr}(r)p_{t}p_{r}-g^{rr}(r)p_{r}^{2}-\frac{1}{r^{2}}\left(p_{\theta}^{2}+\frac{p_{\varphi}^{2}}{\sin^{2}\theta}\right)},
\end{eqnarray}
\begin{eqnarray}
\label{24}
P_{1}=E=-p_{t},
\end{eqnarray}
\begin{eqnarray}
\label{25}
P_{2}=l_{z}=p_{\varphi},
\end{eqnarray}
\begin{eqnarray}
\label{26}
P_{3}=L=\sqrt{p_{\theta}^{2}+\frac{p_{\varphi}^{2}}{\sin^{2}\theta}},
\end{eqnarray}
and a generating function
\begin{eqnarray}
\label{27}
S=\int_{\Gamma}p_{\mu}dx^{\mu}=-Et+l_{z}\varphi+\int_{\Gamma}p_{r}dr+\int_{\Gamma}p_{\theta}d\theta,
\end{eqnarray}
where the integrals are line integrals along the geodesic line $\Gamma$ with constants $E$, $l_{z}$, $L$, and $m$. Obviously all the new momentums are constant. The corresponding conjugate variables are given by
\begin{eqnarray}
\label{28}
Q^{0}=\frac{\partial S}{\partial m}=-m\int_{\Gamma}\frac{dr}{-g^{tr}E+g^{rr}p_{r}},
\end{eqnarray}
\begin{eqnarray}
	\label{29}
	Q^{1}=\frac{\partial S}{\partial E}=-t+\int_{\Gamma}\frac{g^{tt}E-g^{tr}p_{r}}{g^{tr}E-g^{rr}p_{r}}dr,
\end{eqnarray}
\begin{eqnarray}
	\label{30}
	Q^{2}=\frac{\partial S}{\partial l_{z}}=\varphi-l_{z}\int_{\Gamma}\frac{d\theta}{p_{\theta}\sin^{2}\theta},
\end{eqnarray}
\begin{eqnarray}
	\label{31}
	Q^{3}=\frac{\partial S}{\partial l}=-L\int_{\Gamma}\frac{dr}{r^{2}(-g^{tr}E+g^{rr}p_{r})}+L\int_{\Gamma}\frac{d\theta}{p_\theta}.
\end{eqnarray}
Thus, we transform the original coordinates $(t,r,\theta,\varphi,p_{t},p_{r},p_{\theta},p_{\varphi})$ to angle-action variables $(Q^{\mu},P_{\nu})$ in phase space. Keeping in mind that all integrals are line integrals along trajectories with fixed $m$, $E$, $l_{z}$, and $L$. In terms of $(Q^{\mu},P_{\nu})$, the
Hamiltonian reads $H=-P_{0}/2$, and the Vlasov equation is $\partial f/\partial Q^{0}=0$. It was shown in \cite{Rioseco:2016jwc} that for static and spherical configuration the distribution function takes the form
\begin{eqnarray}
\label{32}
f(x^{\mu},p_{\nu})=\mathcal{F}(P_{0},P_{1},P_{3}).
\end{eqnarray}
Considering the Eddington-Finkelstein form of the Schwarzschild-like metric (\ref{14}), we introduce, as done in \cite{Rioseco:2016jwc}, several dimensionless variables $\widetilde{\tau}$, $\xi$, $\pi_{\xi}$, $\pi_{\theta}$, $\varepsilon$, $\lambda$, and $\lambda_{z}$ as

\begin{eqnarray}
	\label{33}
	t=M\widetilde{\tau},~
	r=M\xi,~
	p_{r}=m\pi_{\xi},~
	p_{\theta}=Mm\pi_{\theta},~
	E=m\varepsilon,~
	L=Mm\lambda,~
	l_{z}=Mm\lambda_{z}.
\end{eqnarray}
From $r=M\xi$, we get
\begin{eqnarray}
\label{34}
N=1-\frac{2}{\xi}.
\end{eqnarray}
Let $\eta=1$, then the contravariant metric components in Eq. (\ref{14}) can be written as
\begin{eqnarray}
\label{35}
g^{tt}=-\frac{\frac{l\xi}{\xi-2}+1+\frac{2}{\xi}}{1+l}, \qquad g^{tr}=\frac{2}{(1+l)\xi}, \qquad g^{rr}=\frac{1}{1+l}\left(1-\frac{2}{\xi}\right).
\end{eqnarray}
Using the dimensionless variables to represent the momentum $p_{\theta}$ and $p_{r}$, we have
\begin{eqnarray}
\label{36}
\pi_{\theta}=\pm\sqrt{\lambda^{2}-\frac{\lambda_{z}^{2}}{\sin^{2}{\theta}}}, \qquad \pi_{\xi\pm}=\frac{(1-N\eta)\varepsilon\pm\sqrt{(1+l)\left[\varepsilon^{2}-U_{\lambda}(\xi)\right]}}{N}.
\end{eqnarray}
In terms of dimensionless variables, the effective potential $U_{\lambda}(\xi)$ (\ref{21}) can be written as
\begin{eqnarray}
\label{37}
U_{\lambda}(\xi)=N\left(1+\frac{\lambda^{2}}{\xi^{2}}\right)=\left(1-\frac{2}{\xi}\right)\left(1+\frac{\lambda^{2}}{\xi^{2}}\right).
\end{eqnarray}
In the following sections, we denote $\epsilon(\pi_{\xi\pm})=\pm1$.

\begin{figure}
\centering
\includegraphics[height=4in,width=5in]{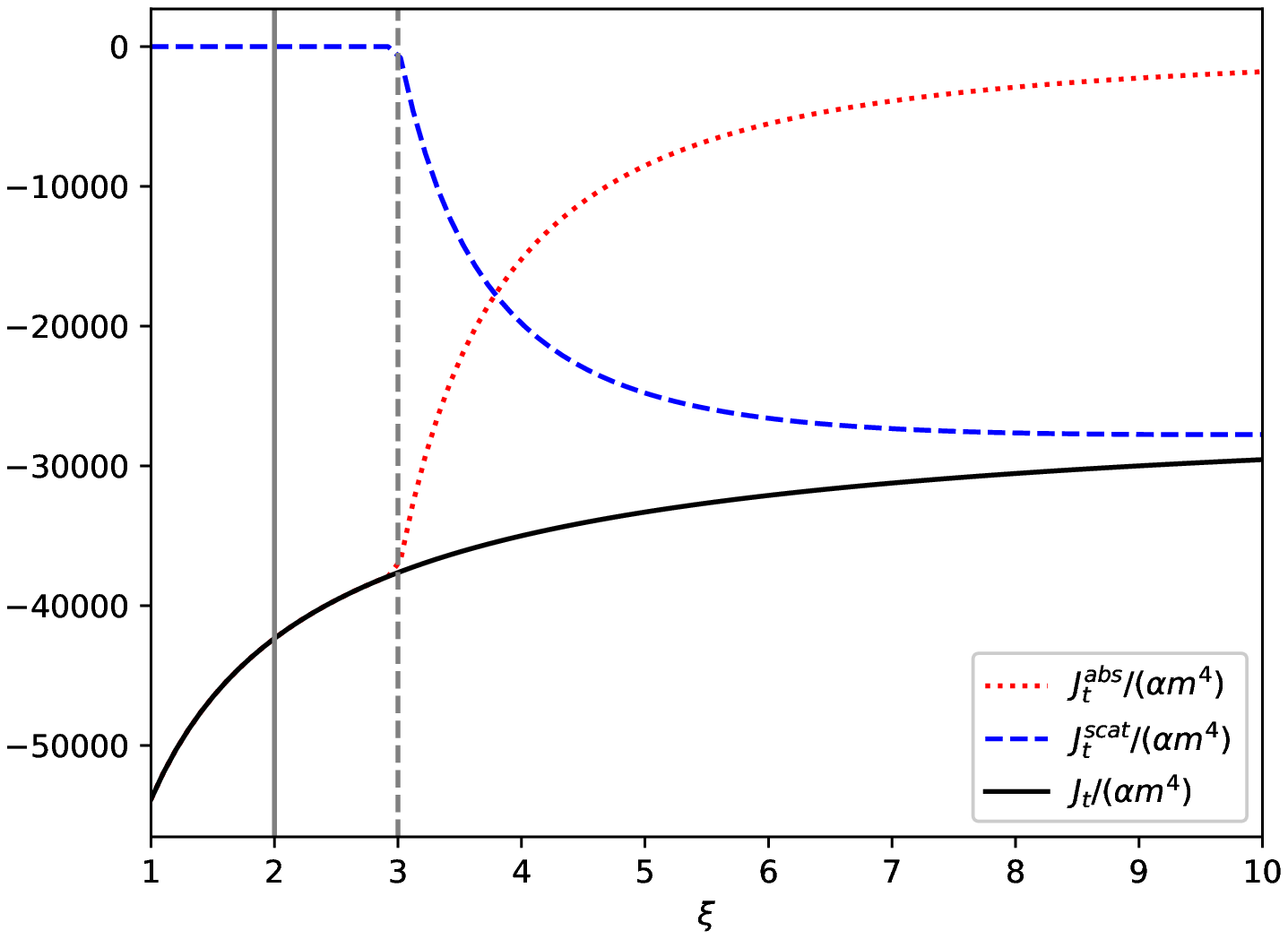}
\caption{Sample graphs of $J_{\rm{t}}^{(\rm{abs})}/(\alpha m^4)$, $J_{\rm{t}}^{(\rm{scat})}/(\alpha m^4)$, and $J_{\rm{t}}/(\alpha m^4)$ with the parameters $\beta=1$. Vertical lines denote the location of the black-hole horizon and the photon sphere.}
\label{fig1}
\end{figure}
\begin{figure}
\centering
\includegraphics[height=4in,width=5in]{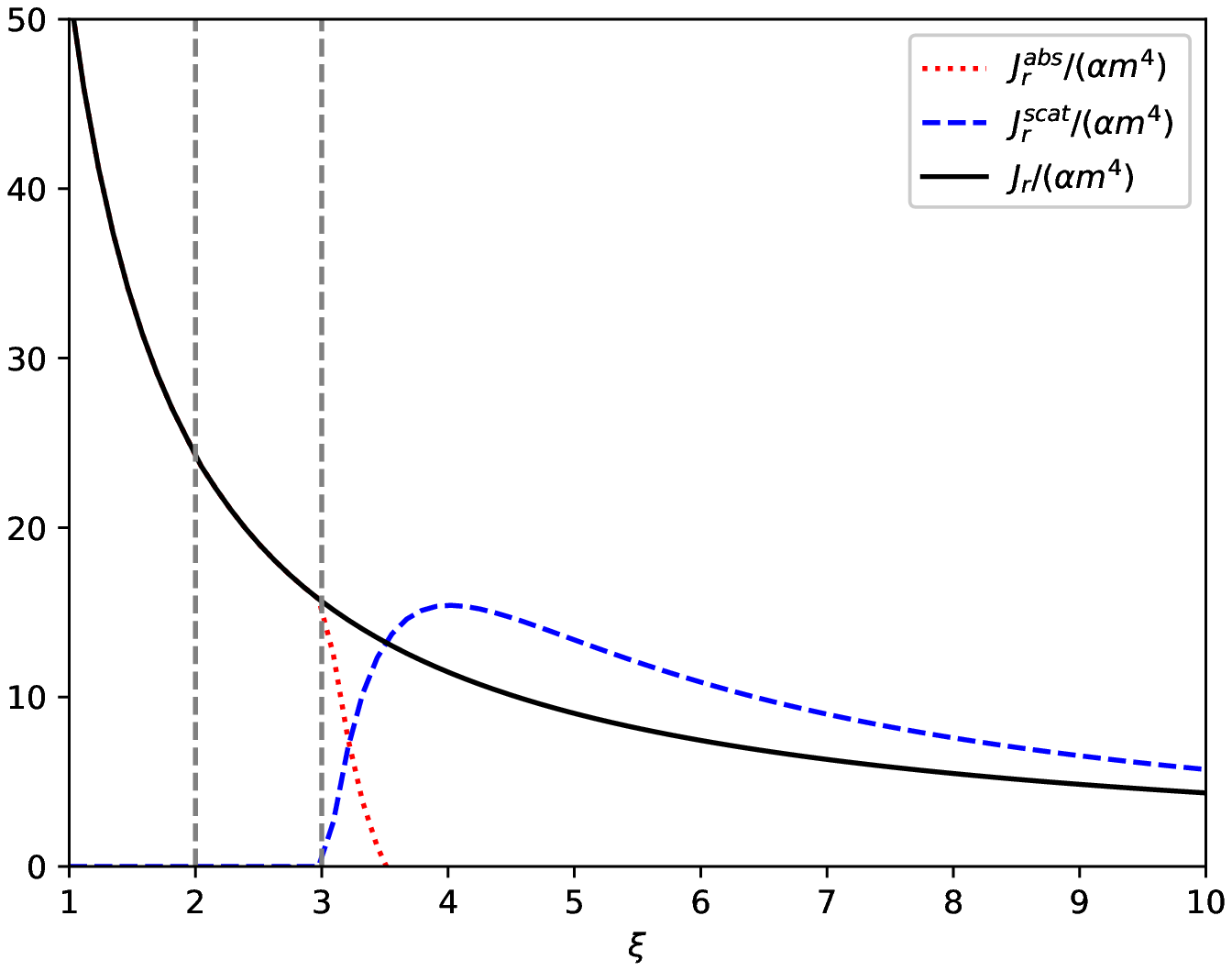}
\caption{Sample graphs of $J_{\rm{r}}^{(\rm{abs})}/(\alpha m^4)$, $J_{\rm{r}}^{(\rm{scat})}/(\alpha m^4)$, and $J_{\rm{r}}/(\alpha m^4)$ with the parameter $\beta=1$ and $l=0.01$. Vertical lines denote the location of the black-hole horizon and the photon sphere.}
\label{fig2}
\end{figure}
\begin{figure}
\centering
\includegraphics[height=4in,width=5in]{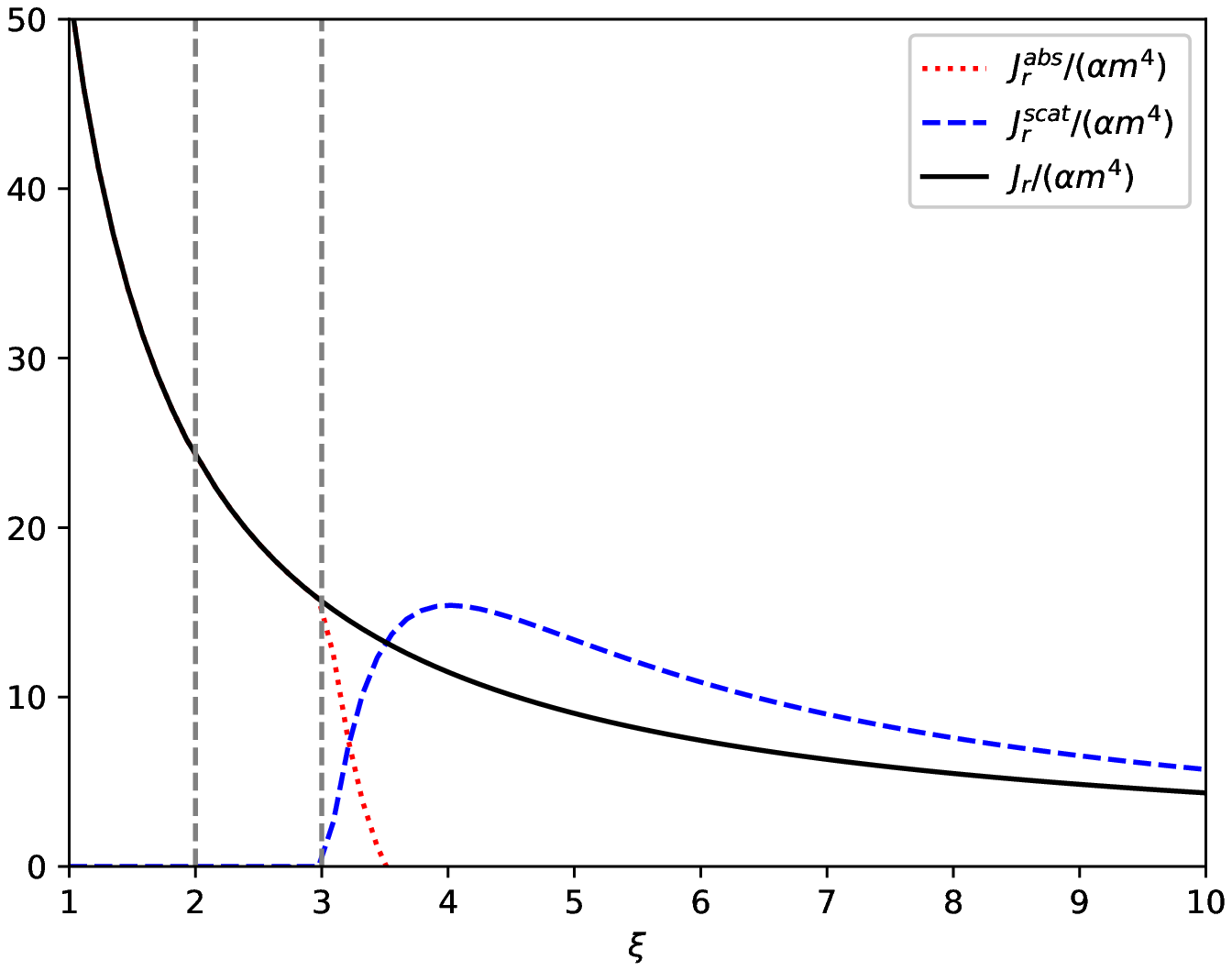}
\caption{Sample graphs of $J_{\rm{r}}^{(\rm{abs})}/(\alpha m^4)$, $J_{\rm{r}}^{(\rm{scat})}/(\alpha m^4)$, and $J_{\rm{r}}/(\alpha m^4)$ with the parameter $\beta=1$ and $l=-0.01$. Vertical lines denote the location of the black-hole horizon and the photon sphere.}
\label{fig3}
\end{figure}

\section{The distribution function}
The key to finding the particle current density is to find a suitable distribution function corresponding to the solution of the Vlasov equation. The distribution function $f$ describing a relativistic and nondegenerate gas in thermal equilibrium in a flat spacetime is called the J{\"u}ttner or Maxwell-J{\"u}ttner distribution which can be written as

\begin{eqnarray}
\label{38}
f(x^{\mu},p_{\nu})=\alpha\delta(\sqrt{-p_{\mu}p^{\mu}}-m)\exp(-\beta\varepsilon),
\end{eqnarray}
where $\alpha$ is a normalization constant, $\varepsilon=E/m$ as defined in (\ref{33}), and $\beta=m/(k_{B}T)$ with $T$ the temperature and $k_{B}$ the Boltzmann constant. The partice density at infinite distances is given by
\begin{eqnarray}
\label{39}
n_{\infty}=4\pi\alpha{m^{4}}\frac{K_{2}(\beta)}{\beta},
\end{eqnarray}
where $K_{2}$ is defined as the modified Bessel function of the second kind \cite{Werner1963}.

\begin{figure}
		\centering
		\includegraphics[height=4in,width=5in]{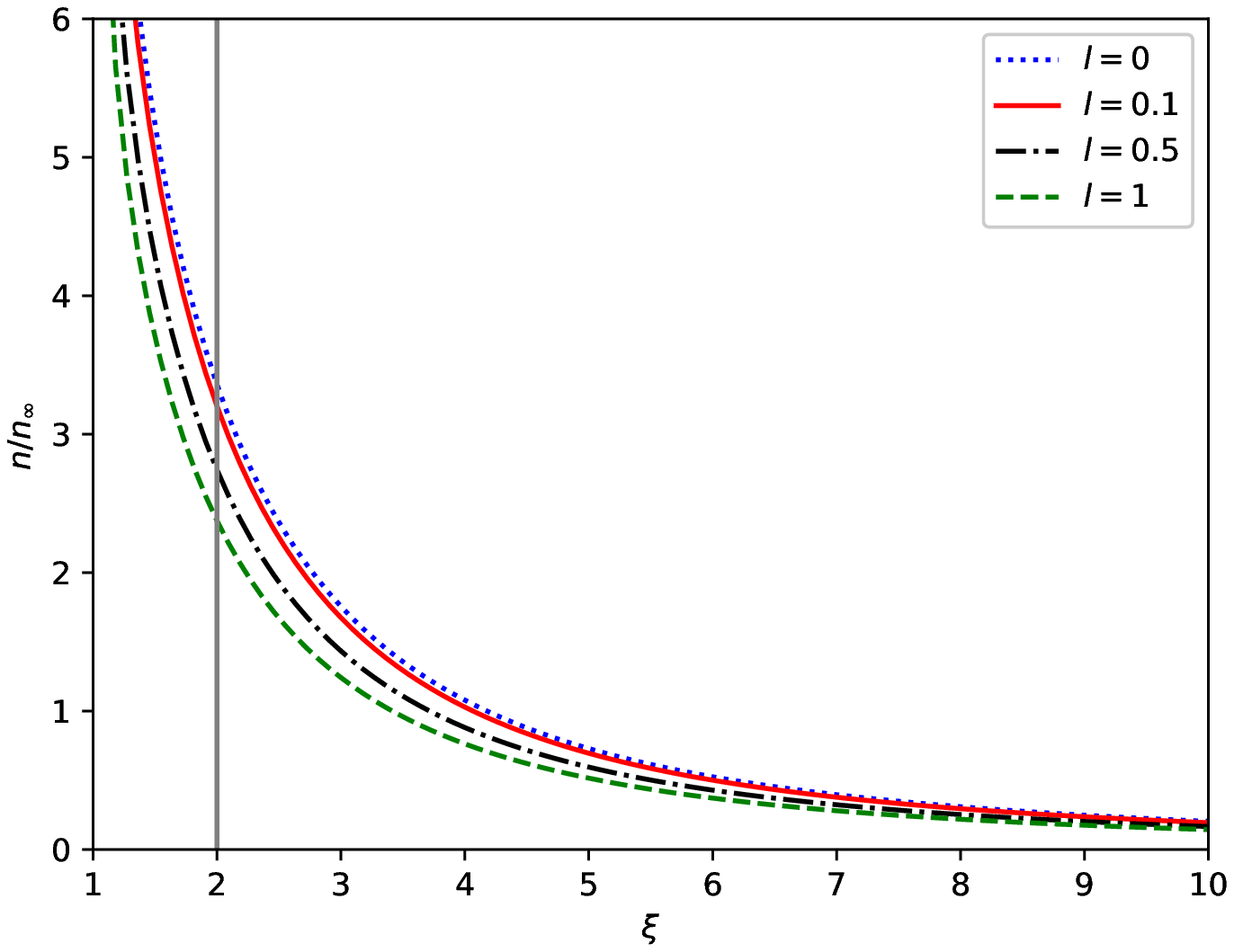}
		\caption{The ratio $n/n_{\infty}$ vs the parameter $\xi$ for $\beta=0.5$. Different graphs correspond to $l=0, 0.1, 0.5$, and $1$. The vertical line marks the location of the black-hole horizon.}
		\label{n1}
\end{figure}
\begin{figure}
		\centering
		\includegraphics[height=4in,width=5in]{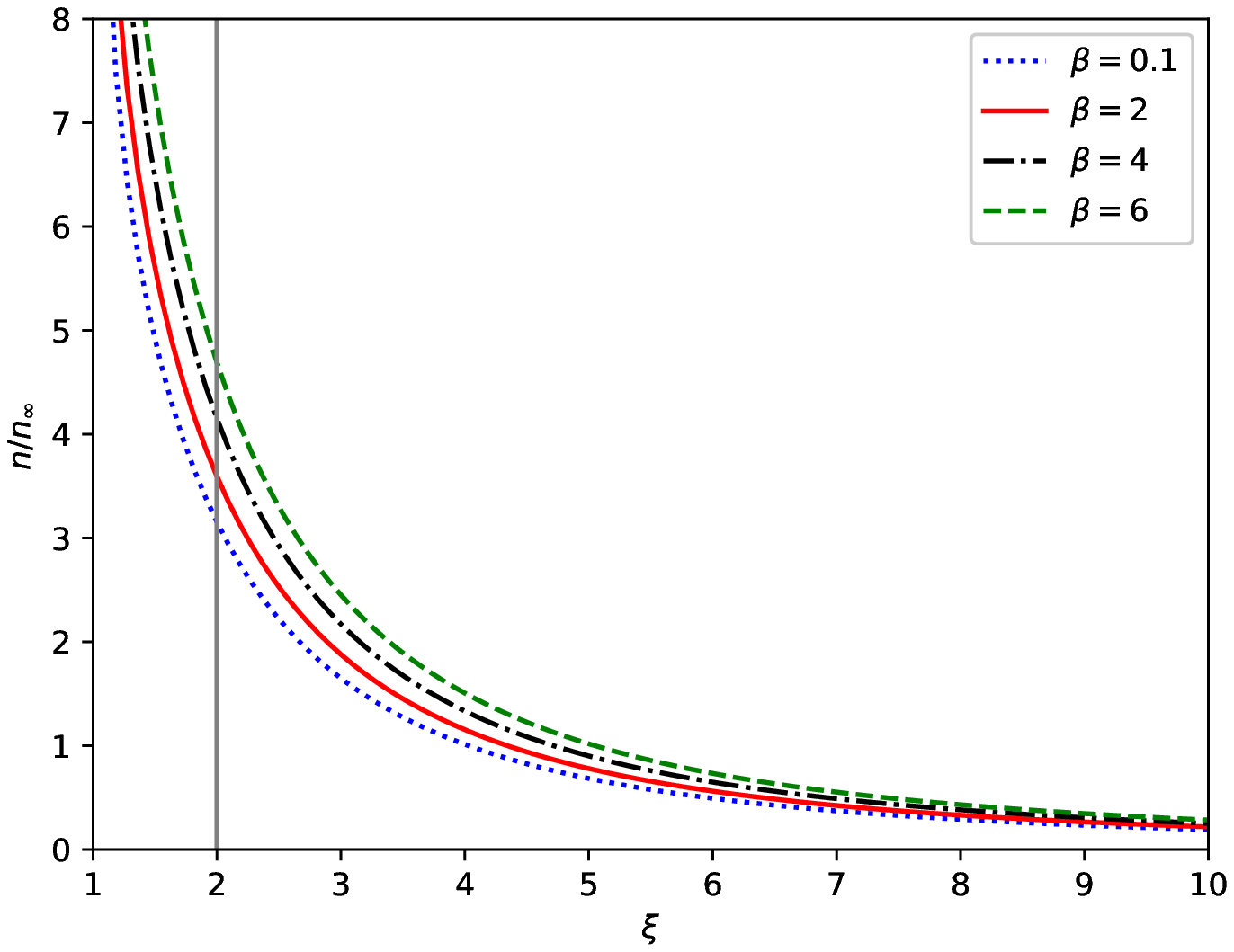}
		\caption{The ratio $n/n_{\infty}$ vs the parameter $\xi$ for $l=0.1$. Different graphs correspond to $\beta=0.1, 2, 4$, and $6$. The vertical line marks the location of the black-hole horizon.}
		\label{n2}
\end{figure}
\begin{figure}
		\centering
		\includegraphics[height=4in,width=5in]{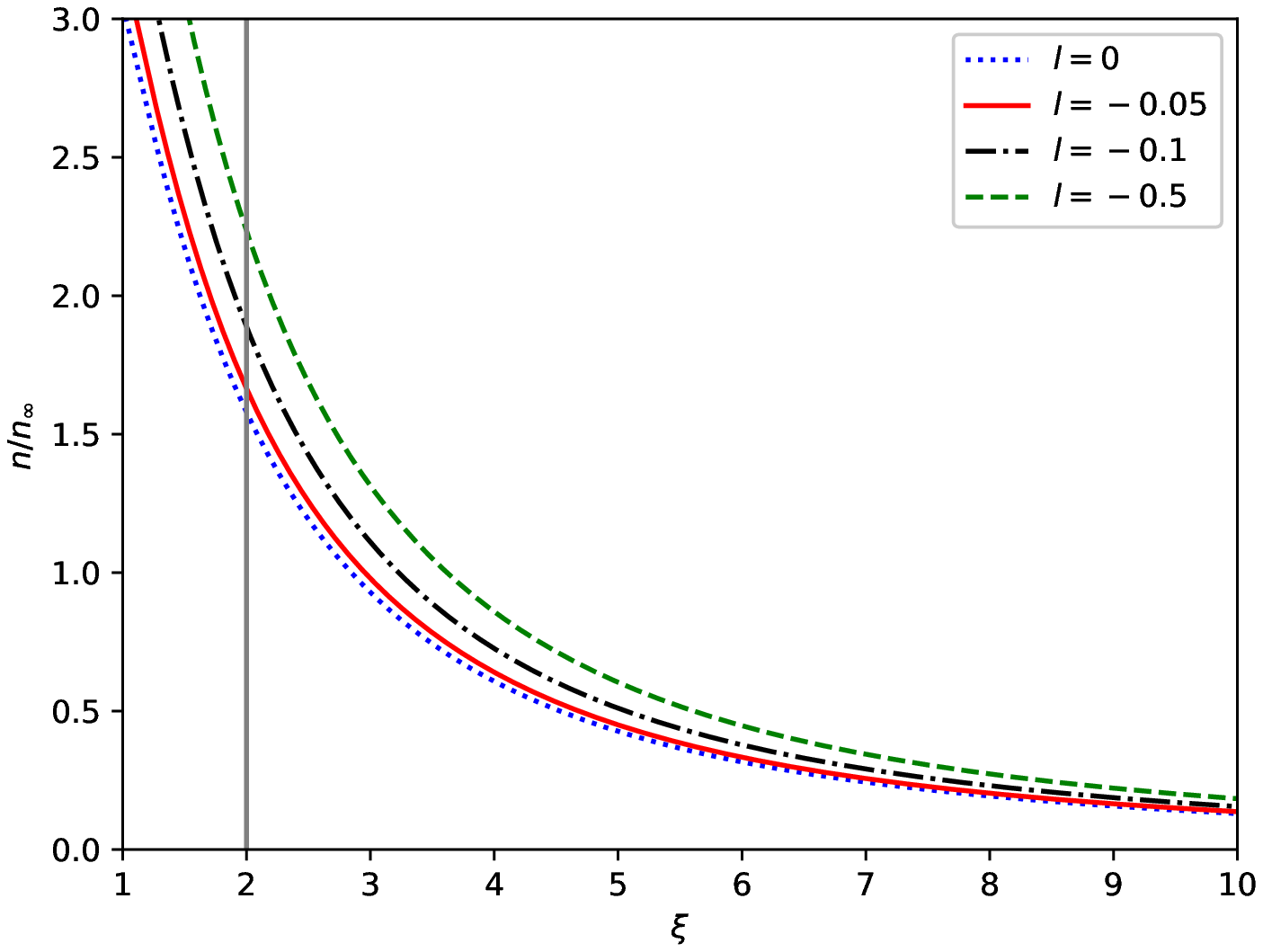}
		\caption{The ratio $n/n_{\infty}$ vs the parameter $\xi$ for $\beta=1$. Different graphs correspond to $l=0, -0.05, -0.1$, and $-0.5$. The vertical line marks the location of the black-hole horizon.}
		\label{n3}
\end{figure}
\begin{figure}
		\centering
		\includegraphics[height=4in,width=5in]{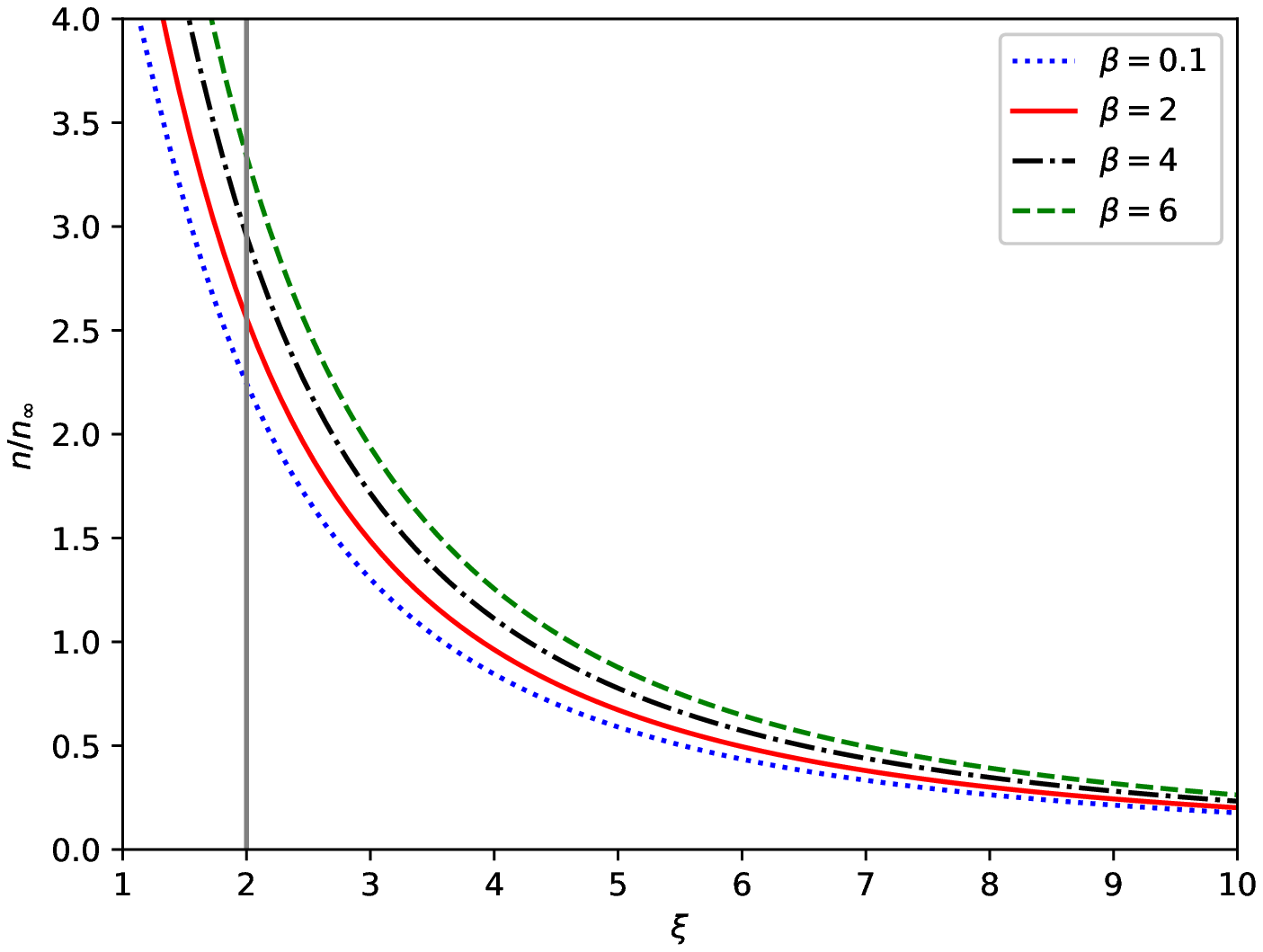}
		\caption{The ratio $n/n_{\infty}$ vs the parameter $\xi$ for $l=-0.1$. Different graphs correspond to $\beta=0.1, 2, 4$, and $6$. The vertical line marks the location of the black-hole horizon.}
		\label{n4}
\end{figure}

\section{Momentum integrals}
In this section, we will discuss the properties of the effective potential for the Schwarzschild-like metric and present the particle current density and accretion rate.
\subsection{Properties of the effective potential}
In order to calculate the integral in the expression of particle current density, we need to know the integral region. The effective potential for the Schwarzschild-like metric (\ref{14}) is
\begin{eqnarray}
	\label{40}
	U_{\lambda}(\xi)=\left(1-\frac{2}{\xi}\right)\left(1+\frac{\lambda^{2}}{\xi^{2}}\right).
\end{eqnarray}
Therefore taking the derivative of $U_{\lambda}$ with respect to $\xi$, we have
\begin{eqnarray}
	\label{41}
	\frac{\mathrm{d}U_{\lambda}}{\mathrm{d}\xi}=\frac{2(\xi^{2}-\lambda^{2}\xi+3\lambda^{2})}{\xi^{4}}.
\end{eqnarray}
Let $\frac{\mathrm{d}U_{\lambda}}{\mathrm{d}\xi}=0$, we obtain the location of the extreme points of $U_{\lambda}$. If $\lambda^{2}>12$, there is a local maximum at
\begin{eqnarray}
	\label{42}
	\xi_{\rm{max}}=\frac{\lambda^{2}}{2}\left(1-\sqrt{1-\frac{12}{\lambda^{2}}}\right),
\end{eqnarray}
and a local minimum at
\begin{eqnarray}
	\label{43}
	\xi_{\rm{min}}=\frac{\lambda^{2}}{2}\left(1+\sqrt{1-\frac{12}{\lambda^{2}}}\right).
\end{eqnarray}
Obviously, the potential $U_{\lambda}(\xi)\to1$ for $\xi\to\infty$. For $\lambda^{2}$ growing from $\lambda^{2}=12$ to infinity, the local maximum $\xi_{\rm{max}}$ decreases from $\xi_{\rm{max}}=6$ to $\xi_{\rm{max}}=3$; while $\xi_{\rm{min}}$ grows from $\xi_{\rm{min}}=6$ to infinity. At the extreme value of $\xi$, the relationship between $\xi$ and $\lambda$ satisfies $\lambda^{2}=\xi^{2}/(\xi-3)$. So at $\xi_{\rm{max}}$ or $\xi_{\rm{min}}$, we have
\begin{eqnarray}
	\label{44}
	U_{\lambda}(\xi)=\left(1-\frac{2}{\xi}\right)\left(1+\frac{1}{\xi-3}\right).
\end{eqnarray}
It follows that for $\lambda^{2}$ from $\lambda^{2}=12$ to infinity, $U_{\lambda}(\xi_{\rm{max}})$ grows from $8/9$ to infinity, while $U_{\lambda}(\xi_{\rm{min}})$ grows from $8/9$ to $1$. One of the important values of the angular momentum $\lambda_{\rm{c}}(\varepsilon)$ is to make $U_{\lambda_{\rm{c}}(\varepsilon)}(\xi_{\rm{max}})$ equal to a given value of $\varepsilon^{2}> 1$, i.e.
\begin{eqnarray}
	\label{45}
	U_{\lambda_{\rm{c}}(\varepsilon)}(\xi_{\rm{max}})=\varepsilon^{2}.
\end{eqnarray}
From this equation we derive the solution for the angular momentum
\begin{eqnarray}
    \label{46}
	\lambda_{\rm{c}}(\varepsilon)=\sqrt{\frac{12}{1-4(\delta-\sqrt{\delta^{2}+\delta})^2}},
\end{eqnarray}
where $\delta=\frac{9}{8}\varepsilon^{2}-1$. There are two types of particles. One type of particles originate at infinity and are absorbed by the black hole. The other type of particles move from infinity with high enough angular momentum and are scattered back to infinity. Particles with angular momentum and energy satisfying the relationship $\varepsilon\geq1$ and $\lambda<\lambda_{\rm{c}}(\varepsilon)$ are absorbed by the black hole. The description of the second type of particles is more complicated. The minimal energy $\varepsilon$ of a scattered particle is given by
\begin{eqnarray}
     \label{47}
	\varepsilon_{\rm{min}}=
	\begin{cases}
		\infty & {\xi\leq3}\\
		\sqrt{\left(1-\frac{2}{\xi}\right)\left(1+\frac{1}{\xi-3}\right)} &  {3<\xi\leq4}\\
		1 & {\xi>4}.
	\end{cases}
\end{eqnarray}
The maximum value of the angular momentum should be the solution to the inequality $\varepsilon^{2}{\geq}U_{\lambda}(\xi)$, i.e.,
\begin{eqnarray}
    \label{48}
	\lambda_{\rm{max}}(\varepsilon,\xi)=\xi\sqrt{\frac{\varepsilon^{2}}{1-\frac{2}{\xi}}-1}.
\end{eqnarray}
Given the minimum value of energy and the maximum value of angular momentum, the particles scattered by the black hole satisfy: $\varepsilon_{\rm{min}}\leq\varepsilon<\infty$ and $\lambda_{\rm{c}}(\varepsilon)<\lambda<\lambda_{\rm{max}}(\varepsilon,\xi)$.

\begin{figure}
    	\centering
    	\includegraphics[height=4in,width=5in]{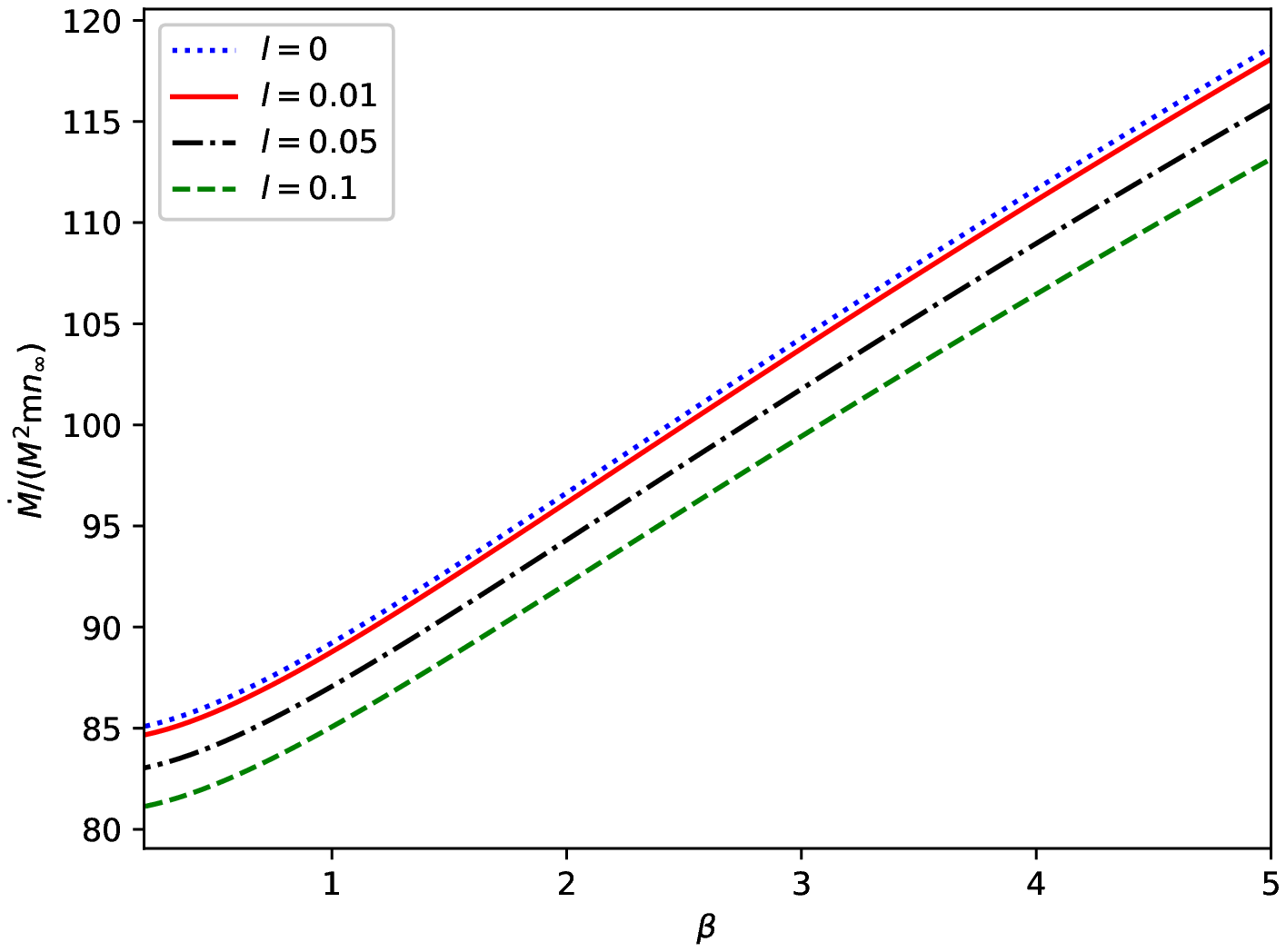}
    	\caption{Mass accretion rate $\dot{M}/(M^{2}mn_{\infty})$ vs the parameter $\beta$. Different graphs correspond to $l=0, 0.01, 0.05, 0.1$.}
    	\label{m1}
\end{figure}
\begin{figure}
    	\centering
    	\includegraphics[height=4in,width=5in]{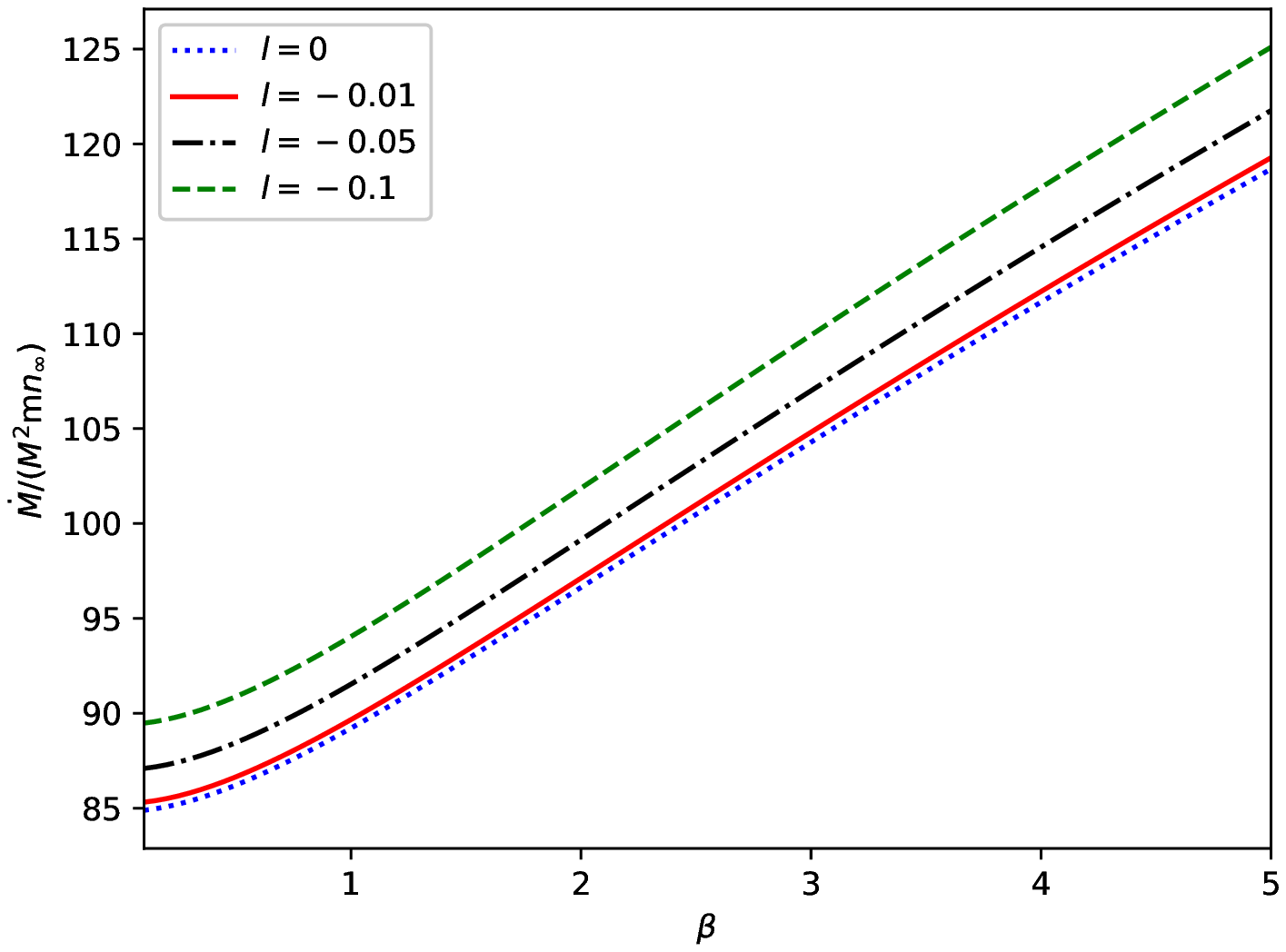}
    	\caption{Mass accretion rate $\dot{M}/(M^{2}mn_{\infty})$ vs the parameter $\beta$. Different graphs correspond to $l=0, -0.01, -0.05, -0.1$.}
    	\label{m2}
\end{figure}

\subsection{The accretion rate}
Now we will calculate the particle current density, to do this, we need to compute the momentum integral in Eq. (\ref{9}). We first introduce a new coordinate $\chi$
\begin{eqnarray}
	\label{49}
	\pi_{\theta}=\lambda\cos\chi, \qquad \lambda_{z}=\lambda\sin\theta\sin\chi,
\end{eqnarray}
with which we can transform the original momentum variables $(p_{t},p_{r},p_{\theta},p_{\varphi})$ to $(\varepsilon,m,\lambda,\chi)$. The original momentum can be represented by using the coordinate $(\varepsilon,m,\lambda,\chi)$, according to
\begin{eqnarray}
	\label{50}
	p_{t}=-m\varepsilon, \qquad p_{\theta}=Mm\lambda\cos\chi, \qquad p_{\varphi}=Mm\lambda\sin\theta\sin\chi.
\end{eqnarray}
These new variables will make it easy to calculate the particle current density. The radial momentum is given by the solution to the following equation
\begin{eqnarray}
	\label{51}
	g^{tt}m^{2}\varepsilon^{2}-2g^{tr}m\varepsilon{p_{r}}+g^{rr}p_{r}^{2}+\frac{m^{2}\lambda^{2}}{\xi^{2}}+m^{2}=0.
\end{eqnarray}
The integral element represented by the new coordinates is
\begin{eqnarray}
	\label{52}
	d^{4}p=\frac{1}{\xi^{2}}\frac{m^{3}\lambda}{\sqrt{\varepsilon^{2}-U_{\lambda}(\xi)}}d{\varepsilon}{dm}{d\lambda}{d\chi}.
\end{eqnarray}
Inserting (\ref{52}) into (\ref{9}), we have the particle current density as
\begin{eqnarray}
	\label{53}
	J_{\mu}(\xi)=\frac{1}{\xi^{2}}\int\frac{p_{\mu}\mathcal{F} m^{3}\lambda}{\sqrt{\varepsilon^{2}-U_{\lambda}(\xi)}}d{\varepsilon}{dm}{d\lambda}{d\chi},
\end{eqnarray}
here we introduce a integral
\begin{eqnarray}
	\label{54}
	\mathcal{F}_{n}=\int_{0}^{\infty}m^{n}\mathcal{F}dm.
\end{eqnarray}
Corresponding to absorbed and scattered particles, the integral quantities $J_{\mu}$ can be divided into two parts, $J_{\mu}=J_{\mu}^{(\rm{abs})}+J_{\mu}^{(\rm{scat})}$. Then we get
\begin{eqnarray}
	\label{55} J_{t}^{(\rm{abs})}=-\frac{\alpha{m^{4}}}{\xi^{2}}\int_{1}^{\infty}d\varepsilon\exp(-\beta\varepsilon)\varepsilon\int_{0}^{\lambda_{c}}d\lambda\int_{0}^{2\pi}d\chi\frac{\lambda}{\sqrt{\varepsilon^{2}-U_{\lambda}(\xi)}},
\end{eqnarray}

\begin{eqnarray}
	\label{56} J_{r}^{(\rm{abs})}=\frac{\alpha{m^{4}}}{\xi^{2}}\int_{1}^{\infty}d\varepsilon\exp(-\beta\varepsilon)\int_{0}^{\lambda_{c}}d\lambda\int_{0}^{2\pi}d\chi\frac{\lambda\pi_{\xi-}}{\sqrt{\varepsilon^{2}-U_{\lambda}(\xi)}},
\end{eqnarray}

\begin{eqnarray}
	\label{57} J_{t}^{(\rm{scat})}=-\frac{2\alpha{m^{4}}}{\xi^{2}}\int_{\varepsilon_{min}}^{\infty}d\varepsilon\exp(-\beta\varepsilon)\varepsilon\int_{\lambda_{c}}^{\lambda_{max}}d\lambda\int_{0}^{2\pi}d\chi\frac{\lambda}{\sqrt{\varepsilon^{2}-U_{\lambda}(\xi)}},
\end{eqnarray}

\begin{eqnarray}
	\label{58} J_{r}^{(\rm{scat})}=\frac{\alpha{m^{4}}}{\xi^{2}}\sum\limits_{\pm}\int_{\varepsilon_{min}}^{\infty}d\varepsilon\exp(-\beta\varepsilon)\int_{\lambda_{c}}^{\lambda_{max}}d\lambda\int_{0}^{2\pi}d\chi\frac{\lambda\pi_{\xi+}}{\sqrt{\varepsilon^{2}-U_{\lambda}(\xi)}}.
\end{eqnarray}
In Figs. (\ref{fig1}), (\ref{fig2}), and (\ref{fig3}), we plot numerically the two components $J_{\rm{t}}$ and $J_{\rm{r}}$, as well as the sum $J_{\mu}=J_{\mu}^{(\rm{abs})}+J_{\mu}^{(\rm{scat})}$ with some special values of the parameters. The components $J_{\mu}^{(\rm{scat})}$ corresponding to the scattered particles vanish for $\xi< 3$, below the photon sphere. The component $J_{\rm{t}}$ is independent of the parameter $l$, while the component $J_{\rm{r}}$ is not. The components $J_{\mu}^{(\rm{abs})}$ and $J_{\mu}^{(\rm{scat})}$ are not smooth, but the total current $J_{\mu}$ is.

The particle density is defined as
\begin{eqnarray}
\label{n}
n=\sqrt{-J_{\mu}J^{\mu}},
\end{eqnarray}
which comes from the relationship between the conserved particle current density and the four-velocity: $J_\mu=nu_\mu$. Sample graphs of the particle density $n$ obtained for different parameters $l$ and for $\beta=0.5$ ($\beta=1$) are shown in Fig. \ref{n1} (\ref{n3}), while for different parameters $\beta$ and for $l=0.1$ ($l=-0.1$) are shown in Fig. \ref{n2} (\ref{n4}). Obviously $n$ is a decreasing function of the radius $\xi$. Increasing $\beta$ or decreasing $l$ will increases the number of particles absorbed into the black hole.

Since $p_{\theta}=Mm\lambda\cos\chi$ and $p_{\varphi}=Mm\lambda\sin\theta\sin\chi$, while $\cos\chi$ and $\sin\chi$ have zero integral values in the interval $(0,2\pi)$, we have $J_{\theta}=0$ and $J_{\varphi}=0$. Therefore $J^{r}=g^{rt}J_{t}+g^{rr}J_{r}$ gives
\begin{eqnarray}
	\label{59}
	J^{r}=\sqrt{\frac{1}{1+l}}\frac{2\pi}{\xi^{2}}\int\epsilon(\pi_{\xi})\mathcal{F}_{4}\lambda{d\varepsilon}{d\lambda}.
\end{eqnarray}
Note that $J_{r}^{(\rm{scat})}=-\frac{g^{tr}}{g^{rr}}J_{t}^{(\rm{scat})}$. The current density of the scattered particle is $0$. So only the current density of the particles absorbed by the black hole contributes to the mass accretion rate which can be easily calculated as
\begin{eqnarray}
	\label{60}
	\dot{M}=-4\pi{m}r^{2}J^{r}=\sqrt{\frac{1}{1+l}}4\pi^{2}M^2m^{5}\alpha\int_{1}^{\infty}\lambda^2_{c}\exp(-\beta\varepsilon)d\varepsilon.
\end{eqnarray}
This integral is generally difficult to calculate, but we can get the analytical expressions for two limiting cases. The mass accretion rate at the low temperature limit ($\beta\to\infty$) is
\begin{eqnarray}
	\label{61}
	\dot{M}(\beta\to\infty)=16\sqrt{\frac{1}{1+l}}M^{2}mn_{\infty}\sqrt{2\pi\beta}.
\end{eqnarray}
While at the high temperature limit ($\beta\to0$), we find
\begin{eqnarray}
	\label{62}
	\dot{M}(\beta\to0)=27\sqrt{\frac{1}{1+l}}{\pi}M^{2}mn_{\infty}.
\end{eqnarray}
From Eqs. (\ref{60}), (\ref{61}), and (\ref{62}), we see that the parameter characterizing the breaking of Lorentz symmetry will reduce the accretion rate, which is consistent with the result for polytropic gas accreted onto a Schwarzschild-like black hole \cite{Yang:2018zef}. In Figs. \ref{m1} and \ref{m2}, we plot numerically $\dot{M}$ vs the parameter $\beta$ for deferent values of the parameter $l$. In all cases the accretion rate $\dot{M}$ increases with the decreasing $l$ or increasing $\beta$.

\section{Conclusions and discussions}
In this paper, we have discussed the spherically symmetric, steady-state accretion problem for Schwarzschild-like black holes. We have obtained the particle current density and the mass accretion rate. We have shown that only the particles absorbed by the black hole contribute to the accretion. We have found that the temporal component of the particle current density $J_{\rm{t}}$ is independent of the parameter $l$, while the radial component $J_{\rm{r}}$ is not. We have presented the concrete expressions for the accretion rate at high and low temperatures. By analyzing the obtained expressions for the accretion rate, we have found a clear dependence of the accretion rate on the model parameters: increasing the parameter $\beta$ or decreasing the parameter $l$ will increases the accretion rate, which is consistent with the result obtained in \cite{Yang:2018zef}. Using Hamiltonian formalism to investigate accretion onto other type black holes deserves further studies.

\begin{acknowledgments}
This work is supported in part by Hebei Provincial Natural Science Foundation of China (Grant No. A2021201034).
\end{acknowledgments}

\bibliographystyle{ieeetr}
\bibliography{acc}

\begin{thebibliography}{10}

\bibitem{Yuan:2014gma}
F.~Yuan and R.~Narayan, ``{Hot Accretion Flows Around Black Holes},'' {\em Ann.
  Rev. Astron. Astrophys.}, vol.~52, pp.~529--588, 2014.

\bibitem{hoyle1939effect}
F.~Hoyle and R.~A. Lyttleton, ``The effect of interstellar matter on climatic
  variation,'' in {\em Mathematical Proceedings of the Cambridge Philosophical
  Society}, vol.~35, pp.~405--415, Cambridge Univ Press, 1939.

\bibitem{1940Obs6339L}
R.~A. {Lyttleton} and F.~{Hoyle}, ``{The evolution of the stars},'' {\em The
  Observatory}, vol.~63, pp.~39--43, Feb. 1940.

\bibitem{Bondi:1944jm}
H.~Bondi and F.~Hoyle, ``{On the mechanism of accretion by stars},'' {\em
  Mon.Not.Roy.Astron.Soc.}, vol.~104, p.~273, 1944.

\bibitem{Bondi:1952ni}
H.~Bondi, ``{On spherically symmetrical accretion},'' {\em
  Mon.Not.Roy.Astron.Soc.}, vol.~112, p.~195, 1952.

\bibitem{michel1972accretion}
F.~C. Michel, ``Accretion of matter by condensed objects,'' {\em Astrophysics
  and Space Science}, vol.~15, no.~1, pp.~153--160, 1972.

\bibitem{begelman1978accretion}
M.~Begelman, ``Accretion of $v> 5/3$ gas by a schwarzschild black hole,'' {\em
  Astronomy and Astrophysics}, vol.~70, p.~583, 1978.

\bibitem{Malec:1999dd}
E.~Malec, ``{Fluid accretion onto a spherical black hole: Relativistic
  description versus Bondi model},'' {\em Phys. Rev. D}, vol.~60, p.~104043,
  1999.

\bibitem{Babichev:2004yx}
E.~Babichev, V.~Dokuchaev, and Y.~Eroshenko, ``{Black hole mass decreasing due
  to phantom energy accretion},'' {\em Phys. Rev. Lett.}, vol.~93, p.~021102,
  2004.

\bibitem{Babichev:2010kj}
E.~Babichev, ``{Galileon accretion},'' {\em Phys. Rev. D}, vol.~83, p.~024008,
  2011.

\bibitem{Rodrigues:2016uor}
M.~E. Rodrigues and E.~L.~B. Junior, ``{Spherical Accretion of Matter by
  Charged Black Holes on f(T) Gravity},'' {\em Astrophys. Space Sci.},
  vol.~363, no.~3, p.~43, 2018.

\bibitem{Contreras:2018gct}
E.~Contreras, A.~Rinc\'on, and J.~M. Ram\'\i{}rez-Velasquez, ``{Relativistic
  dust accretion onto a scale--dependent polytropic black hole},'' {\em Eur.
  Phys. J. C}, vol.~79, no.~1, p.~53, 2019.

\bibitem{Abbas:2018ygc}
G.~Abbas and A.~Ditta, ``{Accretion onto a charged Kiselev black hole},'' {\em
  Mod. Phys. Lett. A}, vol.~33, no.~13, p.~1850070, 2018.

\bibitem{Zheng:2019mem}
J.~Zheng, R.~Ye, J.~Chen, and Y.~Wang, ``{Accretion onto RN-AdS black hole
  surrounded by quintessence},'' {\em Gen. Rel. Grav.}, vol.~51, no.~9, p.~123,
  2019.

\bibitem{Yang:2020bpj}
S.~Yang, C.~Liu, T.~Zhu, L.~Zhao, Q.~Wu, K.~Yang, and M.~Jamil, ``{Spherical
  Accretion Flow onto General Parameterized Spherically Symmetric Black Hole
  Spacetimes},'' {\em Chin. Phys. C}, vol.~45, no.~1, p.~015102, 2021.

\bibitem{UmarFarooq:2020aum}
M.~Umar~Farooq, A.~K. Ahmed, R.-J. Yang, and M.~Jamil, ``{Accretion on high
  derivative asymptotically safe black holes},'' {\em Chin. Phys. C}, vol.~44,
  no.~6, p.~065102, 2020.

\bibitem{Nozari:2020swx}
K.~Nozari, M.~Hajebrahimi, and S.~Saghafi, ``{Quantum Corrections to the
  Accretion onto a Schwarzschild Black Hole in the Background of
  Quintessence},'' {\em Eur. Phys. J. C}, vol.~80, no.~12, p.~1208, 2020.

\bibitem{Panotopoulos:2021ezt}
G.~Panotopoulos, A.~Rincon, and I.~Lopes, ``{Accretion of matter and spectra of
  binary X-ray sources in massive gravity},'' {\em Annals Phys.}, vol.~433,
  p.~168596, 2021.

\bibitem{Iftikhar:2020ykp}
S.~Iftikhar, ``{Accretion onto some singularity-free black holes},'' {\em Int.
  J. Mod. Phys. A}, vol.~35, no.~13, p.~2050062, 2020.

\bibitem{Gao:2008jv}
C.~Gao, X.~Chen, V.~Faraoni, and Y.-G. Shen, ``{Does the mass of a black hole
  decrease due to the accretion of phantom energy},'' {\em Phys.Rev.},
  vol.~D78, p.~024008, 2008.

\bibitem{John:2013bqa}
A.~J. John, S.~G. Ghosh, and S.~D. Maharaj, ``{Accretion onto a higher
  dimensional black hole},'' {\em Phys. Rev. D}, vol.~88, no.~10, p.~104005,
  2013.

\bibitem{Jiao:2016iwp}
L.~Jiao and R.-J. Yang, ``{Accretion onto a Kiselev black hole},'' {\em Eur.
  Phys. J. C}, vol.~77, no.~5, p.~356, 2017.

\bibitem{Ganguly:2014cqa}
A.~Ganguly, S.~G. Ghosh, and S.~D. Maharaj, ``{Accretion onto a black hole in a
  string cloud background},'' {\em Phys. Rev. D}, vol.~90, no.~6, p.~064037,
  2014.

\bibitem{Mach:2013fsa}
P.~Mach and E.~Malec, ``{Stability of relativistic Bondi accretion in
  Schwarzschild-(anti-)de Sitter spacetimes},'' {\em Phys. Rev. D}, vol.~88,
  no.~8, p.~084055, 2013.

\bibitem{Kremer:2020yfg}
G.~M. Kremer and L.~C. Mehret, ``{Post-Newtonian spherically symmetrical
  accretion},'' {\em Phys. Rev. D}, vol.~104, no.~2, p.~024056, 2021.

\bibitem{Tejeda:2019lie}
E.~Tejeda and A.~Aguayo-Ortiz, ``{Relativistic wind accretion on to a
  Schwarzschild black hole},'' {\em Mon. Not. Roy. Astron. Soc.}, vol.~487,
  no.~3, pp.~3607--3617, 2019.

\bibitem{Feng:2022bst}
H.~Feng, M.~Li, G.-R. Liang, and R.-J. Yang, ``{Adiabatic accretion onto black
  holes in Einstein-Maxwell-scalar theory},'' {\em JCAP}, vol.~04, no.~04,
  p.~027, 2022.

\bibitem{Yang:2015sfa}
R.~Yang, ``{Quantum gravity corrections to accretion onto a Schwarzschild black
  hole},'' {\em Phys. Rev. D}, vol.~92, no.~8, p.~084011, 2015.

\bibitem{Yang:2018zef}
R.-J. Yang, H.~Gao, Y.~Zheng, and Q.~Wu, ``{Effects of Lorentz breaking on the
  accretion onto a Schwarzschild-like black hole},'' {\em Commun. Theor.
  Phys.}, vol.~71, no.~5, pp.~568--572, 2019.

\bibitem{Jamil:2008bc}
M.~Jamil, M.~A. Rashid, and A.~Qadir, ``{Charged Black Holes in Phantom
  Cosmology},'' {\em Eur. Phys. J. C}, vol.~58, pp.~325--329, 2008.

\bibitem{Yang:2019qru}
R.~Yang, ``{Constraints from accretion onto a
  Tangherlini\textendash{}Reissner\textendash{}Nordstrom black hole},'' {\em
  Eur. Phys. J. C}, vol.~79, no.~4, p.~367, 2019.

\bibitem{Papadopoulos:1998up}
P.~Papadopoulos and J.~A. Font, ``{Relativistic hydrodynamics around black
  holes and horizon adapted coordinate systems},'' {\em Phys. Rev. D}, vol.~58,
  p.~024005, 1998.

\bibitem{Font:1998sc}
J.~A. Font, J.~M. Ibanez, and P.~Papadopoulos, ``{Nonaxisymmetric relativistic
  Bondi-Hoyle accretion onto a Kerr black hole},'' {\em Mon. Not. Roy. Astron.
  Soc.}, vol.~305, p.~920, 1999.

\bibitem{Zanotti:2011mb}
O.~Zanotti, C.~Roedig, L.~Rezzolla, and L.~Del~Zanna, ``{General relativistic
  radiation hydrodynamics of accretion flows. I: Bondi-Hoyle accretion},'' {\em
  Mon. Not. Roy. Astron. Soc.}, vol.~417, pp.~2899--2915, 2011.

\bibitem{Lora-Clavijo:2015hqa}
F.~D. Lora-Clavijo, A.~Cruz-Osorio, and E.~Moreno~M\'endez, ``{Relativistic
  Bondi\textendash{}Hoyle\textendash{}Lyttleton Accretion Onto a Rotating Black
  Hole: Density Gradients},'' {\em Astrophys. J. Suppl.}, vol.~219, no.~2,
  p.~30, 2015.

\bibitem{Petrich:1988zz}
L.~I. Petrich, S.~L. Shapiro, and S.~A. Teukolsky, ``{Accretion onto a moving
  black hole: An exact solution},'' {\em Phys. Rev. Lett.}, vol.~60,
  pp.~1781--1784, 1988.

\bibitem{Babichev:2008dy}
E.~Babichev, S.~Chernov, V.~Dokuchaev, and {\relax Yu}.~Eroshenko,
  ``{Ultra-hard fluid and scalar field in the Kerr-Newman metric},'' {\em Phys.
  Rev.}, vol.~D78, p.~104027, 2008.

\bibitem{Jiao:2016uiv}
L.~Jiao and R.-J. Yang, ``{Accretion onto a moving Reissner-Nordstr\"om black
  hole},'' {\em JCAP}, vol.~09, p.~023, 2017.

\bibitem{Yang:2021opo}
R.-J. Yang, Y.~Jia, and L.~Jiao, ``{Exact solution for accretion onto a moving
  charged dilaton black hole},'' {\em Eur. Phys. J. C}, vol.~82, p.~502, 2022.

\bibitem{Liu:2009ts}
Y.~Liu and S.~N. Zhang, ``{Exact solutions for shells collapsing towards a
  pre-existing black hole},'' {\em Phys. Lett. B}, vol.~679, pp.~88--94, 2009.

\bibitem{Zhao:2018ani}
S.-X. Zhao and S.-N. Zhang, ``{Exact solutions for spherical gravitational
  collapse around a black hole: the effect of tangential pressure},'' {\em
  Chin. Phys. C}, vol.~42, no.~8, p.~085101, 2018.

\bibitem{Mach:2021zqe}
P.~Mach and A.~Odrzywo, ``{Accretion of Dark Matter onto a Moving Schwarzschild
  Black Hole: An Exact Solution},'' {\em Phys. Rev. Lett.}, vol.~126, no.~10,
  p.~101104, 2021.

\bibitem{Mach:2020wtm}
P.~Mach and A.~Odrzywo\l{}ek, ``{Accretion of the relativistic Vlasov gas onto
  a moving Schwarzschild black hole: Exact solutions},'' {\em Phys. Rev. D},
  vol.~103, no.~2, p.~024044, 2021.

\bibitem{Cieslik:2022wok}
A.~Cie\'slik, P.~Mach, and A.~Odrzywolek, ``{Accretion of the relativistic
  Vlasov gas in the equatorial plane of the Kerr black hole},'' 8 2022.

\bibitem{Rioseco:2016jwc}
P.~Rioseco and O.~Sarbach, ``{Accretion of a relativistic, collisionless
  kinetic gas into a Schwarzschild black hole},'' {\em Class. Quant. Grav.},
  vol.~34, no.~9, p.~095007, 2017.

\bibitem{Cieslik:2020ibk}
A.~Cie\'slik and P.~Mach, ``{Accretion of the Vlasov gas on
  Reissner-Nordstr\"om black holes},'' {\em Phys. Rev. D}, vol.~102, no.~2,
  p.~024032, 2020.

\bibitem{Kostelecky:1989jp}
V.~A. Kostelecky and S.~Samuel, ``{Phenomenological Gravitational Constraints
  on Strings and Higher Dimensional Theories},'' {\em Phys. Rev. Lett.},
  vol.~63, p.~224, 1989.

\bibitem{Kostelecky:2003fs}
V.~A. Kostelecky, ``{Gravity, Lorentz violation, and the standard model},''
  {\em Phys. Rev. D}, vol.~69, p.~105009, 2004.

\bibitem{Bluhm:2004ep}
R.~Bluhm and V.~A. Kostelecky, ``{Spontaneous Lorentz violation,
  Nambu-Goldstone modes, and gravity},'' {\em Phys. Rev. D}, vol.~71,
  p.~065008, 2005.

\bibitem{Casana:2017jkc}
R.~Casana, A.~Cavalcante, F.~P. Poulis, and E.~B. Santos, ``{Exact
  Schwarzschild-like solution in a bumblebee gravity model},'' {\em Phys. Rev.
  D}, vol.~97, no.~10, p.~104001, 2018.

\bibitem{Andreasson:2011ng}
H.~Andreasson, ``{The Einstein-Vlasov System/Kinetic Theory},'' {\em Living
  Rev. Rel.}, vol.~14, p.~4, 2011.

\bibitem{2013Relativistic}
L.~Rezzolla and O.~Zanotti, {\em Relativistic Hydrodynamics}.
\newblock Oxford University Press, 2013.

\bibitem{Werner1963}
W.~{Israel}, ``{Relativistic Kinetic Theory of a Simple Gas},'' {\em Journal of
  Mathematical Physics}, vol.~4, pp.~1163--1181, Sept. 1963.

\end{thebibliography}
\end{document}